\documentstyle[12pt]{article}
\def\appendix#1{
\addtocounter{section}{1}
\setcounter{equation}{0}
\renewcommand{\thesection}{\Alph{section}}
\section*{Appendix \thesection\protect\indent #1}
\addcontentsline{toc}{section}{Appendix \thesection\ \ \ #1}
}

\newcommand{\p}{\partial}

\def\l{\lambda}

\def\n{\nabla}

\def\a{\alpha}
\def\b{\beta}

\def\N{{\cal N}}

\def\be{\begin{equation}}
\def\la{\label}
\def\ee{\end{equation}}
\def\bea{\begin{eqnarray}}
\def\eea{\end{eqnarray}}
\def\e{\varepsilon}
\def\d{\delta}
\def\g{\gamma}
\def\G{\Gamma}

\def\s{\sigma}

\begin{document}
\title{\hfill{UAHEP988} \\
\vspace{1cm}
Quadratic action for type IIB supergravity on $AdS_5\times S^5$}
\author{
G.E.Arutyunov$^{a\,c}$\thanks{arut@genesis.mi.ras.ru}
\mbox{} and S.A.Frolov$^{b\,c}$\thanks{frolov@bama.ua.edu}
\mbox{}
\vspace{0.4cm} \\
$^a$Dipartimento di Matematica, Universita di Milano,
\vspace{-0.1cm} \mbox{} \\
"Federigo Enriques" Via C.Saldini, 50-20133 Milano, Italy
\vspace{0.4cm} 
\mbox{} \\
$^b$Department of Physics and Astronomy, 
\vspace{-0.1cm} \mbox{} \\
University of Alabama, Box 870324,
\vspace{-0.1cm} \mbox{} \\
Tuscaloosa, Alabama 35487-0324, USA
\vspace{0.4cm} 
\mbox{} \\
$^c$Steklov Mathematical Institute,
\vspace{-0.1cm} \mbox{} \\
Gubkin str.8, GSP-1, 117966, Moscow, Russia
\mbox{} 
}
\date {} 
\maketitle 
\begin{abstract}
The quadratic action for physical fields of type IIB supergravity model 
on $AdS_5\times S^5$ is derived starting from the recently found 
covariant action.  All boundary terms that have to be added to the action 
to be used in the AdS/CFT correspondence are determined.
\end{abstract}
\section{Introduction}
It is well-known that the covariant equations  of motion for type IIB 
supergravity \cite{S,SW,HW} can not be derived  from any action because 
of the presence of a self-dual five-form. However,  after eliminating 
some unphysical fields, one arrives at equations  of motion which are not 
manifestly covariant but admit a Lagrangian description.  The formal 
covariance of the Lagrangian can then be provided by  introducing 
auxiliary non-propagating fields. This idea was successfully  applied in 
\cite{ALS,ALT} to construct  a covariant action for type IIB 
supergravity. The covariance  of the action has to be taken with a grain 
of salt since one cannot impose  any covariant gauge conditions on the 
auxiliary fields. Nevertheless, the very existence of the action allows one 
to  study in detail the properties of supergravity. The existence of a 
covariant  action for type IIB supergravity has special interest due to 
the discovery  of the duality between type IIB superstring theory on the 
$AdS_5\times S^5$ background and the four-dimensional  $\N =4$ $SU(N)$ 
super Yang-Mills model \cite{M}. As was argued by Maldacena,  in the 
large $N$ limit and in the limit of large t'Hooft coupling  
$\eta =g_{YM}^2N$  the SYM model may be described by the classical type 
IIB supergravity on $AdS_5\times S^5$.  In particular, the physical 
fields of supergravity  correspond to local primary operators of the SYM 
model.

The conjecture by  Maldacena was further elaborated by Gubser, Klebanov 
and Polyakov \cite{GKP}  and by Witten \cite{W}, who proposed that the 
generating functional of  the connected Green functions in the SYM  model 
coincides with the minimum of the supergravity  action subject to certain 
conditions imposed on supergravity  fields\footnote{The boundary 
conditions of \cite{GKP,W} can be  imposed only on the supergravity 
physical fields that satisfy  second-order differential equations.} at 
the boundary of $AdS_5\times S^5$. It  is worth noting that to make the 
AdS/CFT correspondence complete one has  to add to a supergravity action 
boundary terms. The origin of the boundary  terms was recently clarified 
in \cite{AF1}, where it was shown that they  appear in passing 
from the Hamiltonian formulation of the supergravity to the Lagrangian one.

The AdS/CFT  correspondence has already been used in \cite{AV}-\cite{Y1} 
to compute  some two- and three-point Green functions up to normalization 
constants,  and some preliminary results on four-point Green functions 
have also been obtained \cite{LT, FMMR2,CS}.  However, a detailed 
investigation of the AdS/CFT correspondence  requires the knowledge of 
the type IIB supergravity action. In particular,  to fix the 
normalization constants of two- and three-point  Green functions one has 
to know the quadratic and cubic  actions for physical fields of  
supergravity. To this end one  may try to use the covariant action by 
\cite{ALS,ALT}. In a recent paper  \cite{LMRS} the quadratic action for 
scalar fields corresponding to the chiral primary operators in the $\N =4$ SYM 
was  found by comparing the on-shell values of the covariant action and 
the  most general quadratic action for the scalar fields. This method 
cannot  be used to determine the complete quadratic type IIB supergravity 
action, since for some fields, e.g. fermions,
the on-shell value of the action vanishes,  if one does not take into 
account boundary terms which are in general  unknown. Thus the only way 
to find the quadratic action is to derive it directly from the covariant one.

The aim of the present paper is to determine  the bulk quadratic action 
for physical fields of type IIB supergravity.  Then, by using the 
approach of \cite{AF1} we find all boundary  terms, that have to be added 
to the bulk action in order to get the action 
that has to be used in the AdS/CFT correspondence.

The  plan of the paper is as follows. In Section II we briefly discuss 
the  covariant action of \cite{ALS,ALT} and calculate the quadratic 
covariant  action including all (physical and unphysical) fields of type 
IIB supergravity. In Section III we consider  the part of the action that 
depends on the gravitational fields and the  4-form potential and, in 
particular, reproduce the result of \cite{LMRS}.  However, contrary to 
the results of \cite{LMRS}, there is no nonlocality  in the action. The 
antisymmetric tensor  fields are considered in Section IV, where we 
represent the corresponding  action in the first-order formalism. The 
quadratic action for fermions  is obtained in Section V. In the 
Conclusion we discuss  some possible applications of the results obtained 
and unsolved problems.
The mass spectrum we obtain coincides with the one found in \cite{KRN,GM}.
\section{Covariant action}
\setcounter{equation}{0}
The  covariant action of \cite{ALS,ALT} for type IIB supergravity can be 
written in the form: 
\bea
S&=&\frac{1}{2\kappa^2}\int d^{10}x\sqrt{-g}\left( R - \frac{4}{5!}
F_{M_1\ldots M_5}F^{M_1\ldots M_5}+
\frac{1}{3!}\frac{\partial^La\partial_Ka}{(\partial a)^2}
{\cal F}_{LM_1\ldots M_4}{\cal F}^{KM_1\ldots M_4}\right.\nonumber\\
&+&\left.\frac{2}{5!}\e^{M_1\ldots M_{10}}F_{M_1\ldots M_5}B_{M_6M_7} 
\partial_{M_8} C_{M_9M_{10}}-3{\mbox e}^{-\varphi}(\partial_{[M}B_{NK]})^2
-3{\mbox e}^{\varphi}(\partial_{[M}C_{NK]}- 
\chi\partial_{[M}B_{NK]})^2\right.\nonumber \\
&-&\left.\frac{1}{2}(\partial_{M}\varphi )^2
-\frac{1}{2}{\mbox e}^{2\varphi}(\partial_{M}\chi )^2
 \right) + S(\psi ),
\la{act}
\eea
where $M,N,\ldots ,=0,1,\ldots 9$ and we use the following notations
$$F_{M_1\ldots M_5}=5\partial_{[M_1}A_{M_2\ldots M_5]}+
15(B_{[M_1M_2}\partial_{M_3}C_{M_4M_5]}-C_{[M_1M_2}\partial_{M_3}B_{M_4M_5]});
\quad {\cal F}=F-F^*$$
and all antisymmetrizations are with "weight"1, e.g. $3\partial_{[M}B_{NK]}=
 \partial_{M}B_{NK}-\partial_{N}B_{MK}-\partial_{K}B_{NM}$. The dual 
forms are defined as 
\bea
&&\e_{01\ldots 9}=\sqrt{-g};\quad \e^{01\ldots 9}=-\frac{1}{\sqrt{-g}}
\nonumber\\
&&\e^{M_1\ldots M_{10}}=g^{M_1N_1}\cdots g^{M_{10}N_{10}}\e_{N_1\ldots N_{10}}
\nonumber\\
&& (F^*)_{M_1\ldots M_k}=
\frac{1}{k!}\e_{M_1\ldots M_{10}}F^{M_{k+1}\ldots M_{10}}
 =\frac{1}{k!}\e^{N_1\ldots N_{10}}g_{M_1N_1}\cdots 
g_{M_{k}N_{k}}F_{N_{k+1}\ldots N_{10}}. \nonumber
\eea
$S(\psi )$  denotes the part of the action that depends on fermions. 
Since the  quadratic action for fermions can be easily restored by using 
their equations  of motion, we will discuss in this Section only the 
bosonic part. 

One can easily  verify that (\ref{act}) possesses all the gauge and 
global symmetries  of the conventional covariant equations of motion. 
However, the equations  of motion that follow from (\ref{act}) differ 
from the conventional ones in many aspects.  In particular, the 4-form 
potential now satisfies a second-order differential equation.

The general covariance of the action is achieved  by introducing an 
auxiliary scalar field $a$. As was shown in  \cite{ALS,ALT}, there is a 
gauge symmetry of (\ref{act}) that allows one  to set $a=x_0$. Under this 
choice the action (\ref{act}) does not depend  on the components of the 
4-form potential of the form ${A^{0}}_{MNP}$.

By using the units in which the radius of $S^5$  is set to be unity, the 
$AdS_5\times S^5$ background solution can be written as 
\bea
&&B_2=C_2=\varphi =\chi =0\nonumber\\
&&ds^2=\frac{1}{x_0^2}(dx_0^2+\eta_{ij}dx^idx^j)+d\Omega_5^2=
\dot{g}_{MN}dx^Mdx^N
\nonumber\\
&&R_{abcd}=-\dot{g}_{ac}\dot{g}_{bd}+\dot{g}_{ad}\dot{g}_{bc};
\quad R_{ab}=-4\dot{g}_{ab}\nonumber\\
&&R_{\a\b\g\d}=\dot{g}_{\a\g}\dot{g}_{\b\d}-\dot{g}_{\a\d}\dot{g}_{\b\g};
\quad R_{\a\b}=4\dot{g}_{\a\b}\nonumber\\
&&\bar{F}_{abcde}=\e_{abcde};\quad \bar{F}_{\a\b\g\d\e}=\e_{\a\b\g\d\e},
\la{back}
\eea
 where $a,b,c,\ldots $ and $\a ,\b ,\g ,\ldots$ are the AdS and the 
 sphere indices respectively and $\eta_{ij}$ is the $4$-dimensional 
Minkowski metric.
Representing the gravitational field and the 4-form potential as
$$g_{MN}=\dot{g}_{MN}+h_{MN};\quad A_{MNPQ}=\dot{A}_{MNPQ}+a_{MNPQ},$$
 decomposing (\ref{act}) up to the second order and omitting 
 full-derivative terms, one obtains the quadratic action\footnote{In what 
 follows we omit the common factor $\frac{1}{2\kappa^2}$ in front of the 
action.} 
\bea
S&=&\int d^{10}x\sqrt{-g}\left( -\frac{1}{4}\n_Kh_{MN}\n^Kh^{MN}+
\frac{1}{2}\n_Mh_{KN}\n^Kh^{MN}-\frac{1}{2}\n_Nh_K^K\n_Mh^{MN}\right.
\nonumber\\
&+&\frac{1}{4}\n_Mh_K^K\n^Mh^{N}_N
-\left.\frac{2}{3!}h_N^Mh_L^K\bar{F}_{MKM_1M_2M_3}\bar{F}^{NLM_1M_2M_3}
- \frac{4}{5!}f_{M_1\ldots M_5}f^{M_1\ldots M_5}\right.\nonumber\\
&-& \left.\frac{8}{5!}f_{M_1\ldots M_5}T^{M_1\ldots M_5}
+\frac{1}{3!}\frac{\partial^La\partial_Ka}{(\partial a)^2}
({\cal F}+T)_{LM_1\ldots M_4}({\cal F}+T)^{KM_1\ldots M_4}\right.\nonumber\\
&+&\left.\frac{4}{5!}\e^{M_1\ldots M_{10}}\bar{F}_{M_1\ldots 
M_5}B_{M_6M_7}\partial_{M_8} C_{M_9M_{10}}-3(\partial_{[M}B_{NK]})^2
-3(\partial_{[M}C_{NK]})^2\right.\nonumber \\
&-&\left.\frac{1}{2}(\partial_{M}\varphi )^2
-\frac{1}{2}(\partial_{M}\chi )^2
 \right) + S(\psi ),
\la{qact}
\eea
where
\bea
&&f_{M_1\ldots M_5 } =5\partial_{[M_1}a_{M_2\ldots M_5]};\quad {\cal F} 
=f-f^*\nonumber\\
&&T_{M_1\ldots M_5}=\frac{1}{2}h_K^K\bar{F}_{M_1\ldots M_5}
-5h^K_{[M_1}\bar{F}_{M_2\ldots M_5]K}\nonumber
\eea
One can easily check that the 5-form $T$ is antiself-dual.

The  gauge symmetry of (\ref{qact}) (and of (\ref{act})) allows one to 
impose the following gauge conditions: \bea
&&\n^\a h_{a\a }=0=\n^\a h_{(\a\b )};\quad  
h_{(\a\b )}\equiv h_{\a\b }-\frac{1}{5}\dot{g}_{\a\b}h_\g^\g\la{gh}\\
&&\n^\a a_{M_1M_2M_3\a}=0\la{ga}\\
&&\n^\a B_{M\a }=0=\n^\a C_{M\a}\la{gbc}
\eea
This  gauge choice does not remove all the gauge symmetry of the theory, 
for a detailed discussion of the residual symmetry see \cite{KRN}.

We begin  our study of the action with the most difficult part, 
describing the gravitational fields and the 4-form potential.
\section{Gravitational fields and the 4-form potential}
\setcounter{equation}{0}
In what follows it is convenient to make the change of variable 
$x_0\to {\mbox e}^{x_0}$.  Then the component $\dot{g}_{00}$ of the 
background metric is equal to unity.
In the gauge $a=x_0$ the quadratic action describing the gravitational fields 
and the 4-form potential can be rewritten in the form
\bea
&&S(h,a)=S(h)+S(a)
\la{grff}\\
&&S(h)=\int d^{10}x\sqrt{-g}\left( -\frac{1}{4}\n_Kh_{MN}\n^Kh^{MN}+
\frac{1}{2}\n_Mh_{KN}\n^Kh^{MN}-\frac{1}{2}\n_Nh_K^K\n_Mh^{MN}
\right.\nonumber\\
&&+
\frac{1}{4}\n_Mh_K^K\n^Mh^{N}_N
-\left.\frac{2}{3!}h_N^Mh_L^K\bar{F}_{MKM_1M_2M_3}\bar{F}^{NLM_1M_2M_3}
+\frac{1}{3!}T_{0\mu_1\ldots \mu_4}T^{0\mu_1\ldots \mu_4}\right)\la{gr}\\
&&S(a)=\int d^{10}x\sqrt{-g}\left(
- \frac{4}{5!}f_{M_1\ldots M_5}f^{M_1\ldots M_5}
- \frac{16}{5!}f_{\mu_1\ldots \mu_5}T^{\mu_1\ldots \mu_5}
+\frac{1}{3!}{\cal F}_{0\mu_1\ldots \mu_4}{\cal F}^{0\mu_1\ldots \mu_4}\right)
\la{ff}
\eea
where
$\mu =1,2,\ldots ,9$.

Consider first the part  depending on the 4-form potential. Taking into 
account that the nonvanishing components of $T$ are given by \bea
&&T_{a_1\ldots a_5}=-\frac{1}{2}(h_a^a-h_\a^\a )\e_{a_1\ldots a_5};\quad
T_{\a a_1\ldots a_4}=-h_\a^b \e_{ba_1\ldots a_4}\nonumber\\
&&T_{\a_1\ldots \a_5}=-\frac{1}{2}(h_a^a-h_\a^\a )\e_{\a_1\ldots \a_5};\quad
T_{a \a_1\ldots \a_4}=-h_a^\b \e_{\b\a_1\ldots \a_4}
\la{T}
\eea
we rewrite the action (\ref{ff}) in the form
\bea
S(a)&=&\int d^{10}x\sqrt{-g}\left(
- \frac{1}{2(3!)^2}\e^{0\mu_1\ldots \mu_9}\p_0a_{\mu_1\ldots 
\mu_4}\p_{\mu_5}a_{\mu_6\ldots \mu_9}
-\frac{5}{3}\p_{[\mu_1}a_{\mu_2\ldots \mu_5]}\p^{[\mu_1}a^{\mu_2\ldots \mu_5]}
\right.\nonumber\\
&+&\left.\frac{2}{3}\e^{0ijkl}h_0^\a (\p_\a a_{ijkl}-4\p_la_{ijk\a}) -
\frac{1}{3}(h_a^a-h_\a^\a )\e^{\a_1\ldots\a_5}\p_{\a_5}a_{\a_1\ldots\a_4}\right.
\nonumber\\
&+&\left.\frac{2}{3}\e^{\a\b\g\d\e}h_\e^i 
(\p_i a_{\a\b\g\d}-4\p_\a a_{i\b\g\d}) \right) .
\la{ff1}
\eea
As  was shown in \cite{KRN}, the gauge condition (\ref{ga}) implies that 
the  components of the 4-form potential of the form $a_{i\a\b\g}$ and 
$a_{\a\b\g\d}$ can be represented as follows: \be
a_{i\a\b\g}={\e_{\a\b\g}}^{\d\e}\n_\d\phi_{i\e};\quad
a_{\a\b\g\d}={\e_{\a\b\g\d}}^{\e}\n_\e b
\la{phib}
\ee
It is also convenient to introduce the dual 1- and 0-forms for 
$a_{ijk\a}$ and $a_{ijkl}$:
\be
a_{ijk\a}={\e_{ijk}} ^{l}a_{l\a};\quad
a_{ijkl}=\e_{ijkl}a
\la{aa}
\ee
Then  by using (\ref{phib}) and (\ref{aa}) and the gauge conditions 
(\ref{gh})  and (\ref{ga}), one gets the following expression for the 
action (\ref{ff1}): \bea
S(a)&=&\int d^{10}x\sqrt{-g}\left( -16\p_0b\n_\a^2a +8\n_\a a\n^\a a
 +8\n_i\n_\a^2 b\n^i b-8\n_\a^2 b\n_\b^2 b -8(h_a^a-h_\a^\a ) \n_\b^2b 
\right.\nonumber\\
&+&16\p_0\phi_i^\a (\n_\b^2-4)a_\a^i -8a^{i\a}(\n_\b^2-4)a_{i\a}+
8\n_ia_\a^i\n_ja^{j\a} +16h_0^\a\n_ia^i_\a\nonumber\\
&+&8\n_i\phi_{j\a}(\n_\b^2-4)(\n^i\phi^{j\a}-\n^j\phi^{i\a})-
8(\n_\b^2-4)\phi_{i\a}(\n_\g^2-4)\phi^{i\a}+16h^i_\a (\n_\b^2-4)\phi^{\a}_i
\nonumber\\
&-&\frac{1}{2}\e^{i_1\ldots i_4}\e^{\a_1\ldots\a_5}
\p_0a_{i_1i_2\a_1\a_2}\p_{\a_3}a_{i_3i_4\a_4\a_5} -
2\n_ia_{jk\a\b}(\n^ia^{jk\a\b}-2\n^ja^{ik\a\b})\nonumber\\
&+&\left.
2a_{ij\a\b}(\n_\g^2-6)a^{ij\a\b} \right).
\la{ff2}
\eea
We  see that (\ref{ff2}) is a sum of actions for the scalar fields, the 
vector fields and the antisymmetric tensor fields on $AdS_5$. 

The  action for the antisymmetric fields will be treated first. Although 
the  action is not manifestly covariant with respect to the isometry 
group  of $AdS_5$, one achieves the covariance by introducing the 
additional fields $a_{0i\a\b}$ that obey the follo wing equations:
\bea
a_{0i\a\b}=\frac{1}{4}(-\n_\rho^2+6)^{-1}\e_{ijkl}\e_{\a\b\g\d\e}
\n^\g\n^ja^{kl\d\e}
\nonumber
\eea
Then the  action for the antisymmetric fields can be rewritten in the 
equivalent and manifestly covariant form: \bea
S=\int d^{10}x\sqrt{-g}\left( -
\frac{1}{2}\e^{a_1\ldots a_5}\e^{\a_1\ldots\a_5}
\p_{a_5}a_{a_1a_2\a_1\a_2}\p_{\a_3}a_{a_3a_4\a_4\a_5}
+2a_{ab\a\b}(\n_\g^2-6)a^{ab\a\b} \right).
\la{asf1}
\eea
Now expanding the fields $a_{ab\a\b}$ in terms of the spherical harmonics
\bea
a_{ab\a\b}=\frac{1}{2}a^+_{ab\a\b}+\frac{1}{2}a^-_{ab\a\b},\quad 
a^\pm_{ab\a\b}(x,y)=\sum\, b_{ab}^{I_{10},\pm}(x)Y_{[\a\b]}^{I_{10},\pm}(y),
\nonumber
\eea
where the harmonics are eigenfunctions of the operator
\bea
(^*\n )Y_{[\a\b]}\equiv {\e_{\a\b}}^{\g\d\e}\n_\g Y_{[\d\e]};\quad 
(^*\n )Y_{[\a\b]}^{k,\pm}=\pm 2i(k+2)Y_{[\a\b]}^{k,\pm},\quad k\ge 0,
\nonumber
\eea
we rewrite action (\ref{asf1}) as follows
\bea
S(b_{ab})=-\int_{AdS_5} d^{5}x\sqrt{-g_a}\sum\, (k+2)\left( 
\frac{i}{2}\e^{a_1\ldots a_5}
b_{a_1a_2}^{k,+}\p_{a_3}b_{a_4a_5}^{k,-}
+ (k+2)b_{ab}^{k,+}b^{ab}_{k,-} \right).
\la{asf2}
\eea
Here and in what follows we suppose that the spherical harmonics of all types
are orthonormal.
We see that  action (\ref{asf2}) reproduces the part of the spectrum of 
\cite{KRN,GM} for the antisymmetric fields.

This action  cannot be used in the AdS/CFT correspondence because it's 
on-shell value  vanishes. As follows from \cite{AF2}, one has to add to 
(\ref{asf2}) the boundary term \bea
I(b_{ab})=\int_{\p AdS_5} d^{4}x\sqrt{-\bar{g}}\sum\, \frac{k+2}{2}
b_{ij}^{k,+}b^{ij}_{k,-},
\la{btb}
\eea
where the indices  $i,j$ are contracted with the help of the metric 
induced on the boundary  of $AdS_5$, and $\bar{g}$ is the determinant of 
the induced metric. Then  the sum of the bulk action (\ref{asf2}) and the 
boundary term (\ref{btb}) is the action that  leads to the 
conformally-invariant two-point Green functions \cite{AF2}.

Now consider the part of (\ref{ff2}) that  depends on the scalar fields 
$b$ and $a$. Eliminating the field $a$ by  using it's equation of motion 
one gets \bea
S(b)=\int d^{10}x\sqrt{-g}\left( 
-8\n_a\n_\a b\n^a\n^\a b-8\n_\a^2 b\n_\b^2 b -8(h_a^a-h_\a^\a ) \n_\a^2b \right)
\la{ab1}
\eea
Note that the action is manifestly covariant.

To represent the action for the vector fields  in a covariant form we 
introduce an auxiliary field $\varphi_{0\a}$  which obeys the equation 
$\varphi_{0\a}=\n_ia^i_\a$, and rewrite the action as follows \bea
S(\phi )&=&\int d^{10}x\sqrt{-g}\left( 
16\p_0\phi_i^\a (\n_\b^2-4)a_\a^i -8a^{i\a}(\n_\b^2-4)a_{i\a}+
16\varphi_{0\a}\n_ia^{i\a}-8\varphi_{0\a}\varphi_{0}^{\a}\right.
\nonumber\\
&+&\left. 16h_0^\a\n_ia^i_\a +  
8\n_i\phi_{j\a}(\n_\b^2-4)(\n^i\phi^{j\a}-\n^j\phi^{i\a})\right.\nonumber\\
&-&\left.
8(\n_\b^2-4)\phi_{i\a}(\n_\g^2-4)\phi^{i\a}+16h^i_\a (\n_\b^2-4)\phi^{\a}_i
\right)
\la{aphi1}
\eea
The field $a_{i\a}$ satisfies the following equation of motion
\bea
a_{i\a}=\p_0\phi_{i\a} +(-\n_\b^2+4)^{-1}\n_i(\varphi_{0\a}+h_{0\a}).
\nonumber
\eea
Introducing the field $\phi_{0\a}$ by the formula $(\n_\b^2-4)\phi_{0\a}=
 \varphi_{0\a}+h_{0\a}$ and eliminating the field $a_{i\a}$ from 
(\ref{aphi1}), we get \bea
S(\phi )&=&\int d^{10}x\sqrt{-g}\left(
8\n_a\phi_{b\a}(\n_\b^2-4)(\n^a\phi^{b\a}-\n^b\phi^{a\a})-
8(\n_\b^2-4)\phi_{a\a}(\n_\g^2-4)\phi^{a\a}\right.\nonumber\\
&+&\left. 16h^a_\a (\n_\b^2-4)\phi^{\a}_a -8h_{0\a}h^{0\a}
\right)
\la{aphi2}
\eea
Only  the last term violates the manifest covariance of the action. 
However  we will see in a moment that this term is cancelled by a term 
coming from the action for the gravitational fields. 

To this  end, by using the gauge conditions for the gravitational fields 
and eq.(\ref{T}), we rewrite the action $S(h)$ as follows \bea
S(h)=&&\int d^{10}x\sqrt{-g}\left( -\frac{1}{4}\n_Kh_{ab}\n^Kh^{ab}+
\frac{1}{2}\n_ah^{ab}\n^ch_{cb}-\frac{1}{2}\n_ah_c^c\n_bh^{ba}
\right.\nonumber\\
&&+
\frac{1}{4}\n_Mh_a^a\n^Mh^{b}_b
+\left.\frac{1}{2}h_{ab}h^{ab}+\frac{1}{2}(h_{a}^{a})^2\right.\nonumber\\
&&-\left.\frac{1}{2}\n_ah_\a^\a\n_bh^{ba}+\frac{1}{2}\n_bh_a^a\n^bh^{\a}_\a +
\frac{2}{5}\n_\b h_a^a\n^\b h^{\a}_\a +2h_a^ah^{\a}_\a\right.\nonumber\\
&&+ \frac{1}{5}\n_bh_\a^\a\n^bh^{\b}_\b +\frac{3}{25}\n_\g h_\a^\a\n^\g 
h^{\b}_\b -\frac{13}{5}(h_\a^\a )^2\nonumber\\
&&-\frac{1}{4}\n_Kh_{(\a\b )}\n^Kh^{(\a\b )}
-\frac12 h_{(\a\b )}h^{(\a\b )}\nonumber\\
&&\left. -\frac12\n_ah_{b\a}(\n^ah^{b\a}-\n^bh^{a\a})
-\frac12\n_\b h_{a\a}\n^\b h^{a\a}
-6h^a_\a h^{\a}_a +8h_{0\a}h^{0\a}\right)
\la{gr2}
\eea
So,  we see that although the gauge condition $a=x_0$ violates the 
manifest  covariance of the quadratic action with respect to the action 
of the isometry group of $AdS_5\times S^5$, one can restore it by 
introducing auxiliary fields.

There is  no problem with the fields $h^{(\a\b )}$. Being expanded into 
spherical  harmonics  they directly lead to the corresponding part of the 
spectrum of \cite{KRN,GM}.

Consider  the action for the vector fields that is a sum of (\ref{aphi2}) 
and the last line of (\ref{gr2}): \bea
&&S(vect)=\int d^{10}x\sqrt{-g}\left(
8\n_a\phi_{b\a}(\n_\b^2-4)(\n^a\phi^{b\a}-\n^b\phi^{a\a})-
8(\n_\b^2-4)\phi_{a\a}(\n_\g^2-4)\phi^{a\a}\right.\nonumber\\
&&+\left. 16h^a_\a (\n_\b^2-4)\phi^{\a}_a 
 -\frac12\n_ah_{b\a}(\n^ah^{b\a}-\n^bh^{a\a})
-\frac12\n_\b h_{a\a}\n^\b h^{a\a}
-6h^a_\a h^{\a}_a\right)
\la{avect1}
\eea
Expanding the fields into a set of the spherical harmonics as follows
\bea
&&h_{a\a}(x,y)=\sum\, B_a^{I_5}(x)Y_\a^{I_5}(y);\quad 
\phi_{a\a}(x,y)=\sum\, \phi_a^{I_5}(x)Y_\a^{I_5}(y); \nonumber\\
&&(\n_\b^2-4)Y_\a^k=-(k+1)(k+3)Y_\a^k,
\nonumber
\eea
and making the change of variables \cite{KRN}
\bea
A_a^k=B_a^k -4(k+3)\phi_a^k;\quad C_a^k=B_a^k +4(k+1)\phi_a^k
\nonumber
\eea
we obtain the final action for the vector fields:
\bea
S(vect)=&&\int_{AdS_5} d^{5}x\sqrt{-g_a}\sum\left( 
\frac{k+1}{2(k+2)}\left(-\frac14 (F_{ab}(A^k))^2
-\frac12 (k^2-1)(A_a^k)^2\right)\right.\nonumber\\
&&\left. +\frac{k+3}{2(k+2)}\left(-\frac14 (F_{ab}(C^k))^2
-\frac12 (k+3)(k+5)(C_a^k)^2\right)\right) ,
\la{avect2}
\eea
where $F_{ab}(A)=\p_aA_b-\p_bA_a$. 

Now we  proceed with the most complicated part of the action $S(h,a)$ 
which  depends on the gravitational fields $h_{ab}$ and the scalar fields 
$h_\a^\a$ and $b$: \bea
S(h_{ab},\pi ,b)=&&\int d^{10}x\sqrt{-g}\left( -\frac{1}{4}\n_Kh_{ab}\n^Kh^{ab}+
\frac{1}{2}\n_ah^{ab}\n^ch_{cb}-\frac{1}{2}\n_ah_c^c\n_bh^{ba}
\right.\nonumber\\
&&+
\frac{1}{4}\n_Mh_a^a\n^Mh^{b}_b
+\left.\frac{1}{2}h_{ab}h^{ab}+\frac{1}{2}(h_{a}^{a})^2\right.\nonumber\\
&&-\left.\frac{1}{2}\n_a\pi\n_bh^{ba}+\frac{1}{2}\n_bh_a^a\n^b\pi +
\frac{2}{5}\n_\b h_a^a\n^\b\pi +2h_a^a\pi\right.\nonumber\\
&&+\frac{1}{5}\n_b\pi\n^b\pi +\frac{3}{25}\n_\g\pi\n^\g\pi
-\frac{13}{5}\pi^2\nonumber\\
&&\left.
-8\n_a\n_\a b\n^a\n^\a b-8\n_\a^2 b\n_\b^2 b -8(h_a^a-\pi )\n_\a^2b \right)
\la{ahpib}
\eea
where we denote $h_\a^\a =\pi$, following \cite{KRN}.

First of all we need to remove the mixed terms of the form $\pi h_{ab}$ and $bh_{ab}$, i.e. linear in $h_{ab}$. To this end we make the following shift of the gravitational fields:
\be
h_{ab}=\varphi_{ab} + \frac15 \dot{g}_{ab}\eta +2\n_a\n_b\zeta
\la{redef}
\ee
The  requirement that there is no term linear in $\varphi_{ab}$ fixes 
$\eta$ and $\zeta$ to be \be
\zeta =(-\n_\a^2 +3)^{-1}(\frac15 \pi -6b);\quad 
\eta =-10\zeta -20b -\pi
\la{zeta}
\ee
Then  after straightforward but cumbersome calculations the action 
(\ref{ahpib}) is found to be \bea
S(h_{ab},\pi ,b)=&&S(\varphi_{ab})+S(\pi ,b),\nonumber\\
S(\varphi_{ab}) =&&\int d^{10}x\sqrt{-g}\left( 
-\frac{1}{4}\n_K\varphi_{ab}\n^K\varphi^{ab}+
\frac{1}{2}\n_a\varphi^{ab}\n^c\varphi_{cb}-
\frac{1}{2}\n_a\varphi_c^c\n_b\varphi^{ba}
\right.\nonumber\\
&&+\frac{1}{4}\n_M\varphi_a^a\n^M\varphi^{b}_b
+\left.\frac{1}{2}\varphi_{ab}\varphi^{ab}+
\frac{1}{2}(\varphi_{a}^{a})^2\right)
\la{agr1}\\
S(\pi ,b)=&&\int d^{10}x\sqrt{-g}\left( 4(-\n_\a^2+3)^{-1}(\frac15 \pi -6b)
(\n_a^2-5)(\frac15 \pi -6b)+48(\n_ab)^2\right. \nonumber\\
&&-80(\n_\a b)^2-\frac{2}{25}(\n_a\pi )^2-\frac{2}{25}(\n_\a\pi )^2-
\frac{16}{5}\n_ab\n^a\pi -16\n_\a b\n^\a\pi +240b^2\nonumber\\
&&-4\pi^2 -16b\pi\left.
-8\n_a\n_\a b\n^a\n^\a b-8\n_\a^2 b\n_\b^2 b \right)
\la{abpi}
\eea
The absence of  higher-derivative terms in (\ref{abpi}) is explained by 
the general covariance of (\ref{act}).

Now to get the  final action for the scalar fields we expand them in the 
spherical harmonics as \bea
\pi(x,y)=\sum\, \pi^{I_1}(x)Y^{I_1}(y);\quad 
b(x,y)=\sum\, b^{I_1}(x)Y^{I_1}(y); \quad \n_\a^2Y^k=-k(k+4)Y_\a^k,
\nonumber
\eea
and make the redefinition of the fields \cite{LMRS}
\bea
\pi_k=10ks_k+10(k+4)t_k;\quad b_k=-s_k+t_k\nonumber
\eea
Then, after some algebra we obtain the action
\bea
S(s,t)=&&\int_{AdS_5} d^{5}x\sqrt{-g_a}\sum\left( 
\frac{32k(k-1)(k+2)}{k+1}\left( -\frac12 \n_as_k\n^as_k
-\frac12 k(k-4)s_k^2\right)\right.\nonumber\\
&&\left. +\frac{32(k+2)(k+4)(k+5)}{k+3}
\left( -\frac12 \n_at_k\n^at_k
-\frac12 (k+4)(k+8)t_k^2\right)\right) .
\la{ast}
\eea
Note that the  action for the scalar fields $s_k$ coincides with the one 
found in \cite{LMRS}.  The fact that (\ref{ast}) does not depend on $s_0$ 
and $s_1$ means that these modes are gauge.

Finally we discuss the  action (\ref{agr1}) for the gravitational fields. 
Expanding the fields in  a set of spherical harmonics, we see that the 
zero mode describes a  massless graviton on $AdS_5$. We need to show that 
the massive modes describe traceless  symmetric tensor fields. This can 
be done in two ways. First of all one  can use the equation of 
motion\footnote{This equation is valid  only for the $k\ge 2$ modes. One 
can apply the remaining conformal  diffeomorphism to obtain 
$h_a^a=-\frac35 \pi$ 
\cite{KRN}.} $h_a^a+\frac35 \pi =0$  that enters the complete set of 
equations of motion. Then by using  the equations of motion for $b$ and 
$\pi$ one can easily show that  $\varphi_a^a$ vanishes on shell. The 
second way is to decouple $\varphi_a^a$ from the
 traceless part of $\varphi_{ab}$.  To this end we make the following 
change of variables: \bea
&&\varphi_{ab}=\phi_{(ab)} +  \frac15 \dot{g}_{ab}\phi 
-\frac35\n_a\n_b\n_\a^{-2}\phi \la{ch1}\\
&&\varphi_{a}^a=\phi -\frac35\n_a^2\n_\a^{-2}\phi ,
\la{ch2}
\eea
where $\phi_{(ab)}$ is a traceless symmetric tensor.

\noindent Then the action (\ref{agr1}) acquires the form
\bea
S(\varphi_{ab})=&&S(\phi_{(ab)})+S(\phi ),\nonumber\\
S(\phi_{(ab)})=&&\int d^{10}x\sqrt{-g}\left( -\frac{1}{4}\n_K\phi_{(ab)}\n^K\phi^{(ab)}+
\frac{1}{2}\n_a\phi^{(ab)}\n^c\phi_{(cb)}+
\frac{1}{2}\phi_{(ab)}\phi^{(ab)}\right)
\la{agr2}\\
S(\phi )=&&\int d^{10}x\sqrt{-g}\left( \frac15 (
\phi -\frac35\n_a^2\n_\a^{-2}\phi )(-\n_\a^2+3)\phi \right)
\la{aaph}
\eea
So, we see that (\ref{agr2})  is the action for the traceless massive 
symmetric tensor field that  leads to the same equations of motion and 
the same spectrum as was obtained in \cite{KRN,GM}.

It is obvious that the action (\ref{aaph}) leads to the equation 
\be
\phi -\frac35\n_a^2\n_\a^{-2}\phi =0=\varphi_a^a
\la{aaph1}
\ee
Thus we again conclude that   $\varphi_a^a$ vanishes on shell. Although 
the field $\phi$ satisfies the  second-order equation it does not 
describe any dynamical mode.  The reason is that to make the 
transformation (\ref{ch2}) well-defined we have to impose a
 certain boundary  condition on the field $\phi$, ensuring the 
invertibility of  the operator $1-\frac35\n_a^2\n_\a^{-2}$. Then from 
(\ref{aaph1}) we get that $\phi$ always vanishes on shell.

Thus, we have  completed the discussion of the gravitational fields and 
the 4-form  potential, and now we proceed with the antisymmetric fields 
$B$ and $C$. 
\section{Antisymmetric fields}
\setcounter{equation}{0}
The  action for the antisymmetric tensor fields extracted from 
(\ref{qact}) is given by \bea
S&=&\int d^{10}x\sqrt{-g}\left(
-\n_MB_{NK}(\n^MB^{NK}-\n^NB^{MK}-\n^KB^{NM})\right.\nonumber \\
&-& 
\n_MC_{NK}(\n^MC^{NK}-\n^NC^{MK}-\n^KC^{NM})\nonumber \\
&-&
\left. 4\e^{\a\b\g\d\e}B_{\a\b}\p_\g C_{\d\e}-
4\e^{abcde}B_{ab}\p_c C_{de}\right)
\label{a2}
\eea
Although  the equations of motion obtained from the action coincides with 
the ones  from \cite{KRN}, they are not diagonal, and the fields $B$ and 
$C$ do not  correspond to primary operators of the $\N =4$ SYM model. 
Therefore, the main purpose of this section  is to introduce a proper set 
of fields and to   rewrite the action in  terms of the fields. To this 
end it is convenient to replace the two real fields by one complex field: 
\bea
 A=\sqrt{2}(B+iC),\quad  \bar{A}=\sqrt{2}(B-iC),\quad 
 B=\frac{1}{2\sqrt{2}}(A+\bar{A}),\quad  
C=\frac{1}{2\sqrt{2}i}(A-\bar{A})\nonumber
\eea
Rewriting (\ref{a2}) in terms of $A$ and  $\bar{A}$, one gets, up to 
total derivative terms, \bea
S&=&\int d^{10}x\sqrt{-g}\left( -\frac{1}{2}
\n_M\bar{A}_{NK}(\n^MA^{NK}-\n^NA^{MK}-\n^KA^{NM})\right.\nonumber \\
&+& 
\left. i\e^{\a\b\g\d\e}\bar{A}_{\a\b}\p_\g A_{\d\e}+
i\e^{abcde}\bar{A}_{ab}\p_c A_{de}\right)
\label{a3}
\eea
Taking into account that the fields $A_{MN}$ satisfy the gauge conditions
$\n^\a A_{\a\b}=\n^\a A_{\a b}=0$,
one can rewrite action (\ref{a3}) as follows
\bea
S&=&\int d^{10}x\sqrt{-g}\left( 
-\frac{1}{2}
\left(\n_a\bar{A}_{\a\b}\n^aA^{\a\b}+\n_\g\bar{A}_{\a\b}\n^\g A^{\a\b}+
6\bar{A}_{\a\b}A^{\a\b}\right)+i\e^{\a\b\g\d\e}\bar{A}_{\a\b}\p_\g A_{\d\e}
\right.\nonumber \\
&-&\left(\n_a\bar{A}_{b\a}(\n^aA^{b\a}-\n^bA^{a\a}) +
\n_\b\bar{A}_{a\a}\n^\b A^{a\a}+4\bar{A}_{a\a}A^{a\a}\right)\nonumber\\
&-&\left.\frac{1}{2}
\left(\n_a\bar{A}_{bc}(\n^aA^{bc}-\n^bA^{ac}-\n^cA^{ba})+
\n_\a\bar{A}_{ab}\n^\a A^{ab}\right)+i\e^{abcde}\bar{A}_{ab}\p_cA_{de}\right)
\label{a5}
\eea
It is obvious that the first and second  lines of (\ref{a5}) are just the 
actions for scalar and vector fields  respectively on $AdS_5$. These 
actions directly lead to the spectrum found in \cite{KRN,GM}. 

Let us consider the action describing 
the antisymmetric tensor fields on $AdS_5$:
\bea
S_{as}&=&\int d^{10}x\sqrt{-g}\left( 
-\frac{1}{2}
\left(\n_a\bar{A}_{bc}(\n^aA^{bc}-\n^bA^{ac}-\n^cA^{ba})
+\n_\a\bar{A}_{ab}\n^\a A^{ab}\right)\right.\nonumber \\
&+& 
\left. i\e^{abcde}\bar{A}_{ab}\p_cA_{de}\right)
\label{a6}
\eea
Introducing the auxiliary fields  $P_{ab}$ and $\bar{P}_{ab}$, we rewrite 
the action in the first-order formalism: \bea
S_{as}&=&\int d^{10}x\sqrt{-g}\left( 
-\frac{i}{2}\e^{abcde}\bar{P}_{ab}\p_cA_{de}
+\frac{i}{2}\e^{abcde}P_{ab}\p_c\bar{A}_{de}\right.\nonumber\\
&-&\left. 2\bar{P}_{ab}P^{ab} -\frac{1}{2}\n_\a\bar{A}_{ab}\n^\a A^{ab}
+ i\e^{abcde}\bar{A}_{ab}\p_cA_{de}\right)
\label{a7}
\eea
Changing the variables
\bea
&&A_1=\frac{1}{2}(-\n_\a\n^\a +4)^{\frac{1}{4}}A
+(-\n_\a\n^\a +4)^{-\frac{1}{4}}(P-A)\nonumber\\
&&A_2=\frac{1}{2}(-\n_\a\n^\a +4)^{\frac{1}{4}}\bar{A}
-(-\n_\a\n^\a +4)^{-\frac{1}{4}}(\bar{P}-\bar{A})\nonumber
\eea
one gets the following action
\bea
S_{as}&=&-\int d^{10}x\sqrt{-g}\left( 
\frac{i}{2}\e^{abcde}(\bar{A}_{1ab}\p_cA_{1de}
+\bar{A}_{2ab}\p_cA_{2de})\right.\nonumber\\
&+&\left. (\sqrt{(-\n_\a\n^\a +4)} +2)\bar{A}_{1ab}A_1^{ab} 
+(\sqrt{(-\n_\a\n^\a +4)} -2)\bar{A}_{2ab}A_2^{ab} \right)
\label{a8}
\eea
Taking into account that the eigenvalues of the operator 
$\sqrt{(-\n_\a\n^\a +4)}$ are  equal to $k+2;~k\ge 0$, one obtains the 
spectrum of \cite{KRN,GM}. 
This action has the same form as  the action (\ref{asf2}) for the 
antisymmetric fields coming from  the 4-form potential, and, therefore, 
one has to add to the action the following boundary term \bea
I_{as}=\int_{\p AdS_5\times S_5} d^{9}x\sqrt{-\bar{g}}\left( 
\frac{1}{2}\bar{A}_{1ij}A_{1}^{ij}+
\frac{1}{2}\bar{A}_{2ij}A_{2}^{ij}\right)
\label{bta}
\eea
There is no need to add boundary  terms to the actions obtained in the 
previous section for the scalar,  vector and symmetric tensor fields. The 
actions can be directly used for c omputing two-point Green functions in 
the framework of the AdS/CFT correspondence.    Thus we have completed 
the discussion of the quadratic action for  bosonic fields of type IIB 
supergravity, and now we proceed with the analysis of the fermions.
\section{Fermion fields}
\setcounter{equation}{0}
We begin the consideration of the fermion  fields with the simplest 
spinor case.
The action for the spinor field that leads  to the covariant equations of 
motion \cite{S,SW,HW,KRN} has the form \bea
S=\int d^{10}x\sqrt{-g}\left( \hat{\bar{\lambda}}\G^MD_M\hat{\l} -
\frac{i}{2\cdot 5!}\hat{\bar{\lambda}} \G^{M_1\cdots M_5}F_{M_1\cdots 
M_5}\hat{\l} \right)  \la{al1}
\eea
Here $\hat{\bar{\lambda}} =i\l^*\G^{\hat{1}}$  and $D_M$ is a covariant 
derivative. Recall that $M,N,P\ldots$,  $a,b,c\ldots$ and $\a ,\b ,\g 
\ldots$ are curved ten-dimensional,  $AdS_5$ and $S_5$ indices 
respectively. We denote $\hat{M} ,\hat{N} ,\hat{P}\ldots$,  
$\hat{a} ,\hat{b} ,\hat{c} \ldots$, $\hat{\a},\hat{\b} ,\hat{\g} \ldots$  the 
corresponding  flat indices. We choose the following representation of 
the $\G$-matrices \bea
&&\G^a=\s^1\otimes I_4\otimes \g^a,\quad 
\G^\a=-\s^2\otimes\tau^\a\otimes I_4 \nonumber\\
&&\{\G_{\hat{M}} ,\G_{\hat{N}}\} =2\eta_{\hat{M}\hat{N}}, \quad
\{\g_{\hat{a}} ,\g_{\hat{b}}\} =2\eta_{\hat{a}\hat{b}},\quad 
\{\tau_{\hat{\a}} ,\tau_{\hat{\b}}\} =2\d_{\hat{\a}\hat{\b}}
\nonumber
\eea
In this representation the matrix $\G_{11}$ is equal to
\bea
\G_{11} =\G^{\hat{0}}\cdots \G^{\hat{9}} =
\left(
\begin{array}{cc}
I_{16}& 0\\ 0&  -I_{16}
\end{array}\right)
\nonumber\eea
Since the spinor is right-handed, it has the form
\bea
\hat{\l} =\frac12 (1-\G_{11})\hat{\l}=\left(\begin{array}{c}
0\\ \l
\end{array}\right)\nonumber\eea
Taking into account that
\bea
\g^{\hat{0}}\cdots \g^{\hat{4}}=i\cdot I_4,\quad
\tau^{\hat{5}}\cdots \tau^{\hat{9}}=I_4\nonumber\eea
we find
\bea
\frac{i}{2\cdot 5!}\G^{M_1\cdots M_5}F_{M_1\cdots M_5}=
- \left(
\begin{array}{cc}
0& I_{16}\\ 0& 0
\end{array}\right) =-\s^+\otimes I_{16}
\la{gf}
\eea
and rewrite (\ref{al1}) in the form
\bea
S=\int d^{10}x\sqrt{-g}\, \bar{\lambda}\left( \g^aD_a +i\tau^\a D_\a
+1\right)\l .
\la{al2}
\eea
Here and in what  follows we use the same notation $\g^a$ ($\tau^\a$) for 
the $4\times 4$ matrices and for the $16\times 16$ matrices 
$I_4\otimes\g^a$ ($\tau^\a\otimes I_4$).
Expanding the spinor in a set of the spherical harmonics
\bea
&&\l (x,y)=\sum_{k\ge 0}\left( \l_k^+(x)\Xi_k^+(y)+ \l_k^-(x)\Xi_k^-(y)\right)\nonumber\\
&&\tau^\a D_\a\Xi_k^\pm =m^{I_L}\Xi_k^\pm =\mp i(k+\frac52 )\Xi_k^\pm ,
\nonumber
\eea
we obtain the action for the spinor fields
\bea
S=\int d^{5}x\sqrt{-g_a}\, \sum_{k\ge 0}\left(
\bar{\lambda}_k^+\left( \g^aD_a +k+\frac72\right)\l_k^+
+\bar{\lambda}_k^-\left( \g^aD_a -k-\frac32\right)\l_k^-\right)
\la{al3}
\eea
It is obvious that  the action vanishes on shell. Therefore, in the 
framework of the  AdS/CFT correspondence one has to add a boundary term 
to (\ref{al3}). This boundary term \bea
I=\frac12 \int d^{4}x\sqrt{-\bar{g}}\, \sum_{k\ge 0}\left(
-\bar{\lambda}_k^+\l_k^+
+\bar{\lambda}_k^-\l_k^-\right)
\la{btl}
\eea
was found in \cite{HS}  up to a numerical factor, and the factor was 
fixed in \cite{AF1} by using the Hamiltonian formulation of the action.

Now we proceed with the  gravitino field. The action for the gravitino 
field leading to the covariant equations of motion has the form \bea
S=\int d^{10}x\sqrt{-g}\left( \hat{\bar{\psi}}_M\G^{MNP}D_N\hat{\psi}_P +
\frac{i}{4\cdot 5!}\hat{\bar{\psi}}_M\G^{MNP}\G^{M_1\cdots 
M_5}F_{M_1\cdots M_5}\G_N\hat{\psi}_P \right)  \la{agrn1}
\eea
Taking into account that the gravitino field is left-handed
\bea
\hat{\psi}_M =\frac12 (1+\G_{11})\hat{\psi}_M=\left(\begin{array}{c}
\psi_M\\ 0
\end{array}\right) ,\nonumber\eea
we rewrite (\ref{agrn1}) as follows
\bea
S&=&\int d^{10}x\sqrt{-g}\,\left( \bar{\psi}_a\left( \g^{abc}D_b\psi_c 
-i\g^{ab}\tau^\a D_b\psi_\a \right.
+ i\g^{ab}\tau^\a D_\a\psi_b+
\g^{a}\tau^{\a\b} D_\a\psi_\b -\g^{ab}\psi_b\right)\nonumber\\
&+& \bar{\psi}_\a\left( -i\tau^{\a\b\g}D_\b\psi_\g 
-i\g^{ab}\tau^\a D_a\psi_b 
+\g^{a}\tau^{\a\b}D_\b\psi_a
-\left.
\g^{a}\tau^{\a\b} D_a\psi_\b +\tau^{\a\b}\psi_\b\right)
\right)
\la{agrn2}
\eea
As was shown in  \cite{KRN}, by using local supersymmetries, one can 
choose the gravitino field to be of the form \bea
\psi_\a =\psi_{(\a )} +D_{(\a )}\psi +\tau_\a\eta^+\nonumber\eea
where  
$\tau^{\a }\psi_{(\a )}=\tau^{\a }D_{(\a )}\psi =D^\a\psi_{(\a )}=0$,  
and $\eta^+$ is the Killing spinor, which obeys 
$(D_\a +\frac i2\tau_\a )\eta^+ =0$.
Then (\ref{agrn2}) acquires the form
\bea
S&=&\int d^{10}x\sqrt{-g}\,\left( 
 \bar{\psi}^{(\a )}\left(\g^aD_a\psi_{(\a )} -\psi_{(\a )} 
-i\tau^{\b}D_\b\psi_{(\a )}\right)\right.\la{paa}\\
&+&\bar{\psi}_a\left( \g^{abc}D_b\psi_c 
+ i\g^{ab}\tau^\a D_\a\psi_b-\g^{ab}\psi_b-\g^a(\frac45\hat{D}^2\psi 
+5\psi ) \right)\nonumber\\
&+&\bar{\psi}\tilde{D}_{(\a )}\left( -\g^aD_aD^{(\a )}\psi
-\g^{a}\tau^{\a\b} D_\b\psi_a +D^{(\a )}\psi +i\hat{D}D^{(\a )}\psi
\right)\la{psi22}\\
&+&\bar{\psi}_a^+\left( \g^{abc}D_b\psi_c^+ 
+\frac32\g^{ab}\psi_b^+-5i\g^{ab}D_b\eta^+-10i\g^a\eta^+ \right)\nonumber\\
&+&\left.\bar{\eta}^+\left( -10\eta^+-20\g^aD_a\eta^+ 
-5i\g^{ab}D_a\psi_b^+-10i\g^{a}\psi_a^+ \right)\right)
\la{psi02}
\eea
Here $\hat{D}\equiv\tau^\a D_\a$, 
$\tilde{D}_{(\a )}\equiv D_\a -\frac15 \hat{D}\tau_\a$, $\psi_a^+$  
denotes the lowest mode of the gravitino field that is proportional to a 
Killing spinor, and $\psi_a$ does not contain $\psi_a^+$ in its 
expansion into harmonics.

Expanding the spinor $\psi_{(\a )}$ into a set of the spherical harmonics 
\bea
&&\psi_{(\a )}(x,y)=\sum\,  \psi^{I_T}(x)\Xi_{(\a )}^{I_T}(y)\nonumber\\
&&\g^\b D_\b\Xi_{(\a )}^{I_T} =m^{I_T}\Xi_{(\a )}^{I_T} =\mp i(k+\frac52 
)\Xi_{(\a )}^{I_T} ,\quad k\ge 1, \nonumber
\eea
we represent the action (\ref{paa}) for the spinor fields in the form
\bea
S=\int d^{5}x\sqrt{-g_a}\, \sum\, 
\bar{\psi}^{I_T}\left( \g^aD_a +im^{I_T}-1\right)\psi^{I_T} .
\la{paa1}
\eea
One has to add to the action a boundary term of the form (\ref{btl}).
 
To  show that the action (\ref{psi22}) leads to the spectrum obtained in 
\cite{KRN},  we expand the gravitino field $\psi_a$ and the spinor field 
$\psi$ into harmonics and rewrite (\ref{psi22}) as follows \bea
S&=&\int d^{5}x\sqrt{-g_a}\,\sum\,\left( 
\bar{\psi}_a^{I_L}\left( \g^{abc}D_b\psi_c^{I_L} 
+(im^{I_L}-1)\g^{ab}\psi_b^{I_L} -(\frac45 m_{I_L}^2 +5)\g^a\psi^{I_L} 
\right)\right.\nonumber\\
&-&\left.(\frac45 m_{I_L}^2 +5)\bar{\psi}^{I_L}\left( \g^aD_a\psi^{I_L}
-\g^{a}\psi_a^{I_L}-(1+\frac{3i}{5}m^{I_L})\psi^{I_L}\right)\right)
\la{psi23}
\eea
Performing the shift of the gravitino fields
\bea
\psi_a^{I_L}\to \psi_a^{I_L}+D_a\chi_1^{I_L} +\g_a\chi_2^{I_L}
\nonumber
\eea
and requiring the decoupling of the fields from $\psi^{I_L}$, we find
\bea
&&\chi_1^{I_L}= \frac 35\cdot 
\frac{5+2im^{I_L}}{1+2im^{I_L}}\psi^{I_L}\nonumber\\
&&\chi_2^{I_L}= \frac 15\cdot 
\frac{(1-im^{I_L})(5+2im^{I_L})}{1+2im^{I_L}}\psi^{I_L}\nonumber \eea
Then the action (\ref{psi23}) acquires the form
\bea
S&=&\int d^{5}x\sqrt{-g_a}\,\sum\,\left( 
\bar{\psi}_a^{I_L}\left( \g^{abc}D_b\psi_c^{I_L} 
+(im^{I_L}-1)\g^{ab}\psi_b^{I_L}\right)\right.\la{psi24}\\
&+&\left. 
\frac{2(5+2im^{I_L})(5-2im^{I_L})^2}{25(1+2im^{I_L})}
\bar{\psi}^{I_L}\left( \g^aD_a\psi^{I_L}
+(3+im^{I_L})\psi^{I_L}\right)\right)
\la{psi25}
\eea
The action (\ref{psi24})  is the standard action for the Rarita-Schwinger 
field. To show that the  mode $\g^a\psi_a$ is unphysical one can make the 
change of variables \bea
&&\psi_a^{I_L}=  \varphi_{(a)}^{I_L}+D_a\chi^{I_L} -\frac13 
(im^{I_L}-1)\g_a\chi^{I_L} \nonumber\\
&&\g^a\psi_a^{I_L}= \g^aD_a\chi^{I_L} -\frac53 (im^{I_L}-1)\chi^{I_L}
\nonumber
\eea
and get the following expression for (\ref{psi24})
\bea
S&=&\int d^{5}x\sqrt{-g_a}\,\sum\,\left( 
\bar{\varphi}^{(a)}_{I_L}( \g^{b}D_b\varphi_{(a)}^{I_L} 
-(im^{I_L}-1)\varphi_{(a)}^{I_L})\right.\la{psi26}\\
&+&\left. 
(\frac43 (im^{I_L}-1)^2-3)
\bar{\chi}^{I_L}( \g^aD_a\chi^{I_L}
-\frac53 (im^{I_L}-1) \chi^{I_L})\right)
\la{psi27}
\eea
The same reasoning  as after (\ref{aaph1}) leads to the conclusion that 
the spinor $\chi$ always vanishes on shell. 
We see that the kinetic  term for the gravitino fields is just the sum of 
the kinetic terms for the spinor fields. 
\footnote{Since the derivative  $D_a$ contains the Christoffel symbols, 
(\ref{psi26}) is not a sum of actions for spinor fields. We thank 
O.Rychkov for a discussion of the point.}
Then, one can easily show by  using the results obtained in 
\cite{Cor,Vol} and the Hamiltonian  approach of \cite{AF1}, that the 
boundary term one has to add to (\ref{psi26}) again has the form (\ref{btl}). 

The consideration of the modes  $\psi_a^+$ and
the Killing spinor $\eta^+$ goes the same way. One makes the shift 
\bea
\psi_a^+\to \psi_a^+ +\frac53 i\eta^+
\nonumber
\eea
and  arrives at the action (\ref{psi24}) for the massless gravitino and 
the action \bea
S=\int d^{5}x\sqrt{-g_a}\,  
\frac{40}{3}
\bar{\eta}^+\left( \g^aD_a\eta^+
-\frac{11}{2}\eta^+\right)
\la{psi28}
\eea
for the Killing spinor. 
This  completes our derivation of the quadratic action for type IIB 
supergravity on $AdS_5\times S^5$. 
\section{Conclusion}
An important feature of the quadratic action we obtained in this paper 
is that, although we started from the noncovariant gauge condition, the 
final action possesses manifest invariance with respect to the isometry 
group of the $AdS_5\times S^5$ background. It allows one to expect that 
there may exist a covariant description  of the complete action for type 
IIB supergravity on 
$AdS_5\times S^5$. It is an interesting problem to find such a description. 

As the first step in this  direction one could try to obtain the cubic 
action for supergravity. The solution of the problem would  provide us 
with the knowledge of the relative factors in front of the  two- and 
three-point Green functions computed in the  framework of the AdS/CFT 
correspondence and would allow one to compare  them with the ones found 
in the $\N =4$ SYM. 

The next step is to find the supergravity  action up to the fourth order 
in the physical fields. Then one will be  able to compute the four-point 
Green functions and to verify whether the  logariphmic singularities 
found in \cite{FMMR2} cancel.

Another problem to be solved is to find the  spectrum and the quadratic 
action for the supergravity on the  $AdS_5\times{\cal{E}}_5$ background, 
where 
${\cal{E}}_5$ is an Einstein manifold.  A particularly interesting 
example is 
the manifold ${\cal{E}}_5=T^{1,1}$.  As was shown in \cite{KW} the 
supergravity on this background is  dual to a certain large $N$  $\N =1$ 
superconformal field theory, which  describes a non-trivial infrared 
fixed point of the $SU(N)\times SU(N)$ gauge model.   

\noindent{\bf ACKNOWLEDGMENTS}

G.A. would like to thank Prof.F.Magri for the warm  hospitality at the 
Department of Mathematics of the University of Milan.  S.F. would like to 
thank B.Harms for useful discussions.
The work of G.A. was supported by the Cariplo Foundation for
Scientific Research and in part by the RFBI grant N96-01-00608,
and the work of S.F. was supported by the U.S. Department of Energy under
grant No. DE-FG02-96ER40967 and in part by the RFBI grant N96-01-00551.

\end{document}